# Transportation of Static Magnetic Fields by a practically realizable Magnetic Hose

P.-B. Zhou[1], G.-T. Ma[1], H. Liu[1,2], X.-T. Li[1,2], H. Zhang[1,2], C. Yang[1,2], and C.-Q. Ye[1]

[1] Applied Superconductivity Lab., State Key Lab. of Traction Power, Southwest Jiaotong University, Chengdu, Sichuan 610031, China
[2] School of Electrical Engineering, Southwest Jiaotong University, Chengdu, Sichuan 610031, China

*Abstract*—A practically realizable magnetic hose, constructed by wrapping a ferromagnetic cylinder with alternate superconductor-ferromagnet heterostructure, was developed and its capability to transfer the static magnetic fields, e.g., generated by an Nd-Fe-B magnet, was examined in this letter. A diverse dependence of the transfer efficiency on the diameter of the inner cylinder was found in the magnetic hose demonstrators and the underlying cause was clarified by the finite-element simulations. Transfer efficiency of over 50% in terms of a moderate field has been achieved in the best demonstrator of this study, even with a thin sheet merely having moderate magnetism to embody the ferromagnet in the heterostructure. This work links the theoretically derived model with a physical reality and may also conceive fantastic solutions to form a magnetic circuit with minimum leakage or to create a magnetically shielded space, both of which are deemed promising in most electromagnetic devices.

## I. INTRODUCTION

Manipulation of magnetic field to realize functions such as cloaking [Navau 2009, Sanchez 2011, Narayana 2012, Gömöry 2012] and transferring [Navau 2012, Navau 2014, Prat-Camps 2014] has recent years become one of the research focuses in the scientific community of electromagnetism with the aid of the transformation optics [Ward 1996, Pendry 2012], which supplies a theoretical scheme to arbitrarily control the path of the electromagnetic field, either in the static [Gömöry 2012, Navau 2014] or time-varying [Souc 2013, Gömöry 2015, Solovyov 2015] mode. The resultant electromagnetic materials from the transformation optics to achieve the named functions are magnetically anisotropic in general. The combination of superconductor (SC) and ferromagnet (FM), owning respectively diamagnetic [Araujo-Moreira 2000] and ferromagnetic natures, has an attractive potential to conceive novel magnetic structures with extremely high anisotropy, thereby standing out as a promising solution to realize the theoretically designed metamaterials [Jung 2014]. This SC-FM hybrid has been lately introduced to transfer the static magnetic field—transferred to a long distance is thought to be hard by the conventional way of using FM material alone—with the required magnetic parameters of the SC and FM constituents for a perfect transfer being derived from the transformation optics [Navau 2014]. Though the functional validation of such an innovative concept, which is called as magnetic hose, has been performed by a simplified proof-of-principle experiment [Navau 2014, Anlage 2014], a physical reality of this theoretically derived structure, considering the practical constraints of the related materials, does not seem to have come forth yet. This work is devoted to developing a realistic version of the theoretically derived magnetic hose and experimentally testing its characteristics to transfer the static magnetic fields.

## II. DESCRIPTION OF THE STRUCTURE

The realistic version of the magnetic hose proposed in this letter was made by wrapping a ferromagnetic cylinder with an alternate SC-FM heterostructure (see a sketch in Fig. 1), where the inner cylinder serves as a magnetic core to attract the flux lines from the source (e.g., permanent magnet in Fig. 1(a)) and the outer SC-FM heterostructure confines the flux lines axially to suppress the leakage through the lateral surface. The use of a magnetic core in the present version may degrade the transfer efficiency of such a magnetic device in theory, it will however certainly avoid the damage of the superconductivity of the SC constituent restricted by a limited mechanical tolerance [Obradors 2014, Barth 2015], which is one of the critical issues for applications and was idealized in the primary work [Navau 2014].

The magnetic hose was embodied by using the coated superconductor from SuperPower with non-magnetic substrate (SCS4050; 4 mm in width and 0.1 mm in thickness) as the outer SC constituent and a ferromagnetic sheet of 30 µm in thickness (stainless steel 1Cr17) as the inner FM constituent in the SC-FM heterostructure, with the surrounded magnetic core being made of a commercial soft iron. The mechanical condition upon the coated superconductor is further facilitated in practice by building the SC-FM heterostructure in such a way that the coated superconductor therein was placed parallel to the axis of the inner cylinder (see the related insets in Fig. 2), implying that we formed a polygon instead of a circle while winding the SC-FM

Corresponding author: G.-T. Ma (gtma@swjtu.cn).

heterostructure, as clearly demonstrated by the inserts in Figs. 2 and 3. Compared to the bending way adopted by Navau et al. [2014], this way allows the diameter of the inner cylinder to be a minimum incapable of bending the coated superconductor without damage in the superconductivity.

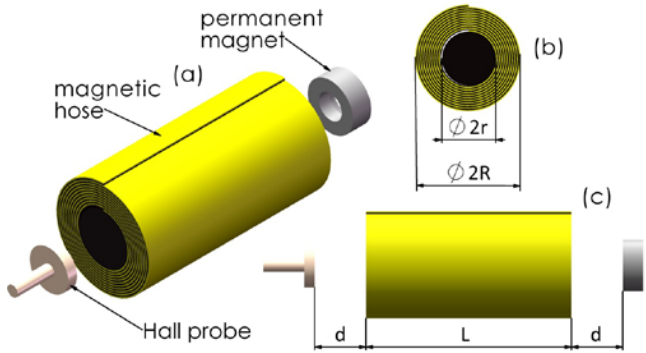

Fig. 1. Sketch of the experimental scenario of the proposed magnetic hose. (a) Three-dimensional view of the magnetic hose used to transfer the static magnetic field of a hollow permanent magnet. (b) The front view of the magnetic hose with the superconducting and the ferromagnetic constituent represented by respectively yellow and black. (c) The side view of the experimental scenario, being the magnetic hose placed coaxially relative to the magnetic source.

## III. RESULTS AND DISCUSSION

Before being cut to pieces with a desired length $L$, the coated superconductor was tested in liquid nitrogen via a four-probe method and the derived critical current is around 104 A with a criterion of electric field $E_c = 10^{-4}$ V/m, which is highly close to the nominal value (100 A) released by the provider and thus confirms the superconductivity of the samples used in the following experiments.

Fig. 1 sketches the experimental scenario of the studied magnetic hose, with a hollow permanent magnet placed at one end as a representative source to generate the magnetic field and a Hall probe (Lakeshore HGCT-3020) deployed at the other end coaxially to measure the transferred magnetic field. A superconducting sheet, having identical geometry to the ferromagnetic sheet, was used to draw the SC-FM heterostructure for simplification. The magnetic field excited by the hollow permanent magnet is 4.8 mT (28 mT) at a distance of $2d(d)$, and the transferred magnetic field by the magnetic hose at a distance of $2d+L$ for all cases is over 1.5 mT, rendering the influence of the geomagnetic field as well as the residual magnetism negligible in our measurements. The SC constituent in the magnetic hose suffered a zero-field-cooling condition in all experiments as we mainly exploited its diamagnetic response in the studied situation. In the default case, the studied magnetic hose has a dimension of $2R = 20$ mm and $L = 40$ mm, whilst its distance to the permanent magnet $d = 10$ mm.

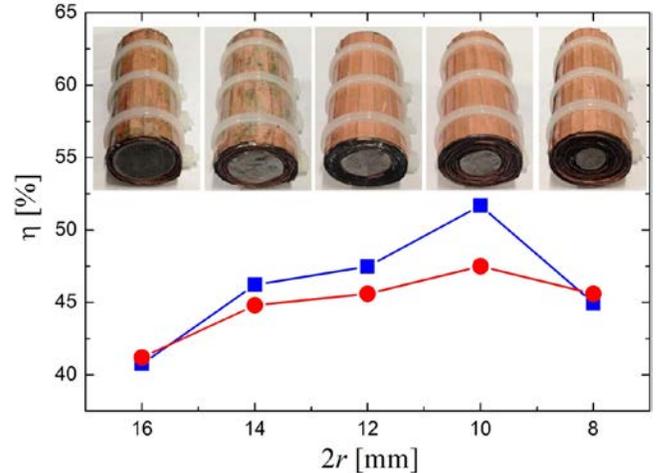

FIG. 2. Measured (solid rectangle) and simulated (solid circle) transfer efficiency of the magnetic hose as a function of the diameter 2r of the ferromagnetic cylinder with 2R = 20 mm and L = 40 mm unvaried (see Fig. 1). The insets are the experimental demonstrators of the magnetic hose with a varied diameter of the inner cylinder from 8 to 16 mm at an interval of 2 mm.

We follows the definition of transfer efficiency by Navau et al. [2014] to appraise the capability of the proposed magnetic hose, viz., $\eta = B_{2d+L}/B_{2d}$, where $B_{2d}$ represents the magnetic field of the magnetic source at a distance of $2d$ and $B_{2d+L}$ is the transferred magnetic field due to the presence of the magnetic hose at a distance of $2d+L$ (see a sketch of the dimension in Fig. 1 (c)).

Varying the diameter ($2r$) of the magnetic core from 8 to 16 mm at an interval of 2 mm, we made a set of magnetic hoses with identical dimension of outer diameter ($2R$) and length ($L$). The measured transfer efficiency of these magnetic hoses as a function of the diameter of the magnetic core was plotted in Fig. 2. This plot displays that, the transfer efficiency is improved considerably from a minimum of 40.8% to a maximum of 51.7% as the decrease of the value of $2r$ from 16 to 10 mm, and then degraded to be 45% with further shrinking the inner cylinder to be $2r = 8$ mm, revealing a diverse dependence of the transfer efficiency on the diameter of the magnetic core. Here worthy of statement is that, the relevant existing predictions by means of idealized numerical simulations (cf. figure S4 in the Supplementary Material by Navau et al. [2014]) could not be used to interpret this testing observation because the physical model in that experimental work is not a practical version of the numerical work, or more concretely, the numerical work simulated a bilayer SC-FM hybrid, whereas this work measured a set of multilayer SC-FM hybrids, with a ferromagnetic core being wrapped by an alternate SC-FM heterostructure. The existence of an optimum portion of the FM constituent in the bilayer SC-FM hybrid to achieve the best transfer seems to be expected, as the extremes of respectively pure SC and baked FM at $r = 0$ and $r = R$ should suggest a maximum in between to entail the introduction of the SC-FM hybrid. However, the case of this experimental work is different because the demonstrator at $r = 0$ would become a fully alternate SC-FM heterostructure with the magnetic core absent, whose transfer efficiency is certainly not zero, but a finite value being smaller than the optimum one.

Understanding the increase in the transfer efficiency observed in Fig. 2 is straightforward because the augment of the surrounding SC-FM heterostructure brings an approach of the proposed magnetic hose to its theoretical version [Navau 2014]. By contrast, the degradation of the transfer function with the magnetic core less than 10 mm in diameter seems to be inconsistent with the theoretical analysis. This divergence is likely to be attributed to (i) even the portion of the SC-FM heterostructure augments, the distortion of the SC polygon to a circle, arising from the way to form the magnetic hose, is aggravated as the decrease of the inner diameter (2r) of the SC-FM heterostructure, which causes more gaps in the vicinity of the magnetic core and consequently, may incur an appreciable leakage of the flux lines through the lateral surface of the magnetic core; (ii) an optimum matching between the magnetic source and the magnetic core naturally exists, being supported in this study by such an evidence that the outer diameter of the hollow Nd-Fe-B magnet is exactly identical to the diameter of the magnetic core, i.e., 10 mm, where the highest transfer efficiency is achieved.

To clarify the above-discussed phenomenon discovered by experiments, we further carried out finite-element simulations by resorting to the prevailing COMSOL Multi-physics software and the related results were plotted in Fig. 2 for comparison. In the simulations, we empirically assigned a relative permeability of respectively 100 and 10 to the magnetic core and the FM constituent in the heterostructure, whereas a value of 0.05 to the SC constituent, which behaves as a diamagnetic material in this study. As the thickness of the SC and FM constituents in the heterostructure is extremely thin, especially for the SC constituent whose SC layer is merely 1 µm, numerical simulation considering real geometry of such a complex object was found to be impossible in an acceptable time. We therefore referred to a simplified version, in which the SC and FM constituents were supposed to be alternated rings with an identical thickness of 0.25 mm. The magnetization of the hollow Nd-Fe-B magnet, finite-element modeled in actual shape and dimensions, was determined by matching the simulated magnetic field with its measured counterpart.

It can be seen from Fig. 2 that, the variation of the transfer efficiency with the diameter of the magnetic core, found in the measured plot, was reproduced by calculation, though the discrepancy in magnitude clearly exists. As we made a set of hypothesis in electromagnetically modeling the magnetic hose, including the geometry as well as the material parameter, the existence of the discrepancy between the measurement and calculation is logically acceptable. More simulations demonstrate that, the general trait of reaching a maximum of transfer efficiency at $2r$ = 10 mm stands independent of the relative permeability assigned to the ferromagnetic and superconducting elements, probably confirming the experimentally observed matching between the magnetic source and the magnetic core.

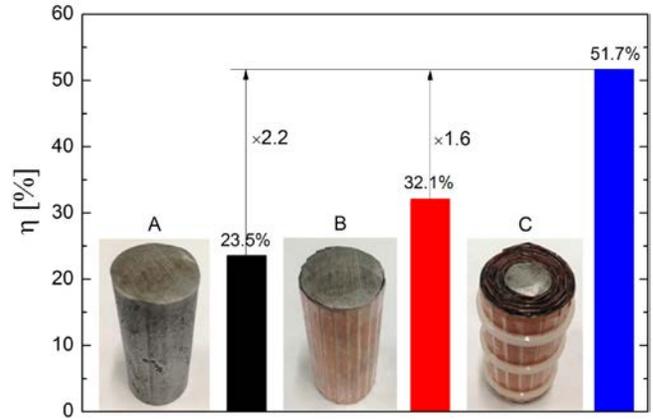

FIG. 3. Illustration of the considerably improved transfer efficiency of the magnetic hose from a naked ferromagnetic cylinder (case A) to a ferromagnetic cylinder wrapped by one superconductor layer (case B) to a ferromagnetic cylinder wrapped by an alternate superconductor-ferromagnet heterostructure (case C).

To check the transfer performance of the proposed magnetic hose, we further carried out experiments on a ferromagnetic cylinder barely or wrapped by one SC layer, having a dimension of 20 mm in diameter and 40 mm in length. As demonstrated in Fig. 3, a considerable improvement in the transfer efficiency has been attained from the naked ferromagnetic cylinder ($\eta$ = 23.5%) to its upgraded version wrapped by one SC layer ($\eta$ = 32.1%) and to the proposed magnetic hose ($\eta$ = 51.7%). This comparison firmly proves the excellent performance of the proposed magnetic hose to transfer the magnetic field in a moderate range—an advancement of practical interest over the low field studies [Navau 2014].

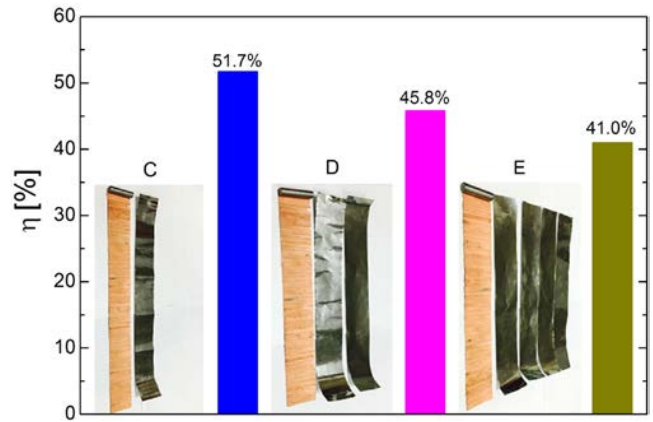

FIG. 4. Dependence of the transfer efficiency of the magnetic hose on the thickness of the ferromagnetic constituent in the alternate superconductor-ferromagnet heterostructure to wrap the ferromagnetic cylinder ($2r$ = 10 mm). As shown in the insets, the thickness of the ferromagnetic constituent increases from one layer (case C) to two layers (case D) to four layers (case E).

As the diamagnetic response of the SC is not strictly reciprocal to the ferromagnetic response of the FM, investigations to find the optimum portion of the SC/FM constituent in the heterostructure were performed and the experimental results were illustrated in Fig. 4. This figure tells that, the increase in the thickness of the FM constituent will degrade the transfer efficiency dramatically in the studied magnetic hose, which suggests that the transfer efficiency of the magnetic hose may be further enhanced by using a thinner ferromagnetic sheet, i.e., less than 30 μm.

Additionally, a longer magnetic hose with $2R$ = 16 mm and $L$ = 60 mm was made to verify the excellent performance of the proposed version using its counterpart by Navau et al. [2014]. At a distance of $d$ = 10 mm, a transfer efficiency as high as 38.2% was achieved by the magnetic hose with an inner cylinder of $2r$ = 10 mm, being significantly higher than the one (20.5%) reported by Navau et al. [2014], of which the SC-FM structure was subject to a weaker field (0.92 vs. 4.8 mT) and a shorter distance $d$ (5 vs. 10 mm).

Given the material property and operating temperature, the transfer efficiency of the proposed magnetic hose could be upgraded by easily shortening the width of the SC constituent to approach a circle in a higher degree, or by tightly coupling the SC and FM constituents via a compatible adhesive to reduce the gap in between, or by directly developing a coated superconductor with the SC ingredient deposited on a high-permeability substrate to form an inherent SC-FM heterostructure.

## IV. CONCLUSION

We have realized a magnetic hose for transferring the static magnetic fields by wrapping a magnetic core with an alternate SC-FM heterostructure from a practical point of view. In addition to transfer magnetic field, the proposed magnetic hose, grounding on its superior capability to regulate the flux lines, can be exploited to form a magnetic circuit with minimum leakage and to create a magnetically shielded space, which are deemed promising properties for a wide range of applications in the electromagnetic devices. Moreover, a cloaking function against incident magnetic field could be intuitively conceived in the proposed magnetic hose by removing the magnetic core and reconstructing the heterostructure with SC being inner and FM being outer.

## ACKNOWLEDGMENT

We thank Z.-G. Deng, J. Zheng, D.-B. He, R. Wang and H. Chen for helpful assistance in experiments. This work was supported in part by the National Natural Science Foundation of China under Grant 51475389, and by the Fundamental Research Funds for the Central Universities under Grant 2682014CX039, and by the Self-determined Projects of the State Key Laboratory of Traction Power under Grants 2013TPL_T05 and 2015TPL_T05.